\documentclass[amssymb,aps,twocolumn,showpacs, superscriptaddress]{revtex4}
\usepackage[]{graphicx}
\usepackage[]{amsmath}
\usepackage{hyperref}

\def\ket#1{| #1 \rangle}

\def\cG{\mathcal{G}}

\def\cS{\mathcal{S}}
\def\cT{\mathcal{T}}

\def\cZ{\mathcal{Z}}

\newtheorem{theorem}{Theorem}
\newtheorem{definition}{Definition}

\def\eq#1{Eq.~\eqref{eq:#1}}
\def\fig#1{Fig.~\ref{fig:#1}}
\def\be{\begin{equation}}
\def\ee{\end{equation}}
\def\BE{\begin{equation}}
\def\EE{\end{equation}}
\newcommand\BF{\begin{figure}[t!]}
\newcommand\EF[2]{\caption{#1}\label{#2}\end{figure}}

\begin{document}

\title{Universal topological phase of 2D stabilizer codes}
\author{H\'ector Bomb\'in}
\affiliation{Perimeter Institute for Theoretical Physics, 31 Caroline St. N., Waterloo, ON, N2L 2Y5, Canada}
\author{Guillaume Duclos-Cianci}
\affiliation{D\'epartement de Physique, Universit\'e de Sherbrooke, Sherbrooke, Qu\'ebec, J1K 2R1, Canada}
\author{David Poulin}
\affiliation{D\'epartement de Physique, Universit\'e de Sherbrooke, Sherbrooke, Qu\'ebec, J1K 2R1, Canada}

\date{\today}

\begin{abstract}
Two topological phases are equivalent if they are connected by a local unitary transformation. 
In this sense, classifying topological phases amounts to classifying long-range entanglement patterns.
We show that all 2D topological stabilizer codes are equivalent to several copies of one universal phase: Kitaev's topological code. 
Error correction benefits from the corresponding local mappings.
\end{abstract}


\maketitle

The theory of Ginzburg and Landau has had a tremendous success at classifying the different phases of matter in terms of local order parameters and spontaneously broken symmetries. However, it fails to classify certain states of nature, such as the different fractional quantum Hall fluids which all have the same local symmetries. The Hamiltonian of these systems has a constant energy gap, and the ground state degeneracy depends on the topology of the space. Crucially, all ground states are locally identical, which explains the failure of the Ginzburg-Landau paradigm.  Instead, the classification of these systems requires the concept of topological order. 

Because topological order reflects the long-scale many-body correlations of the system, it cannot be modified locally. This robustness \cite{BHM10a} is indeed one of the many features that makes topologically ordered systems interesting for quantum information processing \cite{NSSF08a}. It also suggests a natural classification of topological phases: systems that only differ by a local rearrangement of their degrees of freedom belong to the same topological phase. In other words, the different phases are characterized only by their long-range entanglement patterns \cite{CGW10a}.

Another, more conventional, description of these phases is in terms of adiabatic connections. If two local and gapped Hamiltonians are connected by a family of local and gapped Hamiltonians, then it should be possible to adiabatically interpolate between the two without encountering a phase transition. The two systems should therefore be in the same phase. This adiabatic evolution will generate a local unitary transformation \cite{HW05a}, so consequently the two systems will be in the same topological phase according to the definition adopted above. 

Quantum error-correcting codes \cite{KL97a} are intimately related to topological order. To protect the information from local errors, information is encoded into the long-range entanglement of the system. 
A stabilizer code \cite{CRSS98a} is a special type of quantum code that can be defined as the degenerate ground state of a Hamiltonian on $N$ qubits of the form 
\be
H = -\sum_a S_a \quad{\rm with}\quad [S_a,S_b] = 0 \quad \forall a,b
\label{eq:H}
\ee
where the stabilizer operators $S_a$ are Hermitian elements of the Pauli group, i.e. they are constructed from tensor products of the three Pauli matrices $\sigma_x$, $\sigma_y$, and $\sigma_z$ and the identity operator $I$. Stabilizer codes are also {\em frustration free}, meaning that the  $S_a$ do not generate $-1$ under multiplication, so the ground states of $H$ are $+1$ eigenstates of all stabilizers, i.e., $S_a \ket\psi = +\ket\psi$ for all $a$. The $S_a$ form an Abelian group under multiplication, the stabilizer group $\cS$.  When the qubits are embedded on a regular lattice, the code---or its associated Hamiltonian---is said to be {\em local} if each operator $S_a$ has support
on a region of constant size, independent of the system size.
The support of an operator contains those qubits on which it acts nontrivially.

In this Letter, we are interested in stabilizer codes that (i) are local and translationally invariant (LTI), and (ii) are topological, in the sense that no local operator can recover any encoded information---i.e., they have a macroscopic minimum distance in terms of error correction.
If we place our stabilizer in an infinite lattice, this can be formalized as follows.
\begin{definition}
A topological stabilizer code (TSC) is a LTI stabilizer $\cS$ such that
$\cZ(\cS)\propto \cS.$
\end{definition} 
The symbol $\cZ(\cS)$ denotes the centralizer of $\cS$, the group of 
Pauli operators (with bounded support) that commute with all the elements of $\cS$.
Our main result is that the topological phase of any 2D TSC is uniquely determined by its total quantum dimension $\kappa$, or equivalently by its topological entanglement entropy  $S_{\rm topo} = \kappa \log 2$ \cite{KP06a,LW05a}. 
This follows from the existence of a local mapping to $\kappa/2$ copies of Kitaev's topological code (KTC) \cite{Kit03a,DKLP02a}.
We also adapt the result to a class of subsystem stabilizer codes \cite{Pou05b,Bac05a}.

Many considerations motivate this line of research. Firstly, stabilizer codes provide simple models to study many-body quantum physics because they often admit exact solutions, and at the same time can exhibit complex phenomena such as topological order and anyonic excitations \cite{Kit03a,DKLP02a,BM07a}. To our knowledge, this is the first example where the definition of topological order based on  local equivalence \cite{CGW10a} can be directly applied to a class of models in a rigorous manner. Secondly, in the context of error correction, the local equivalence to KTC enables us to directly extend a number of properties of this code to all 2D TSCs. For instance, thermal instability \cite{NO08a,AFH09a}, code tradeoffs \cite{BT08a}, logical operator geometry \cite{KC08a}, and scale invariance \cite{AV08a} all become trivial corollaries of our mapping. In addition, our mapping provides a method to decode any 2D TSC code, while only a handful of special cases previously had solutions \cite{DKLP02a,DP10a,SBT10a}. Thirdly, the local mapping can be used to change encoding during a quantum computation. Because the mapping is local, this change will not propagate errors and is therefore fault-tolerant. This allows to put together the features of different codes---such as having transversal Clifford gates \cite{BM07a}, lower weight stabilizer generators \cite{Kit03a,DKLP02a,B10b1}, etc.---and suggests a natural generalization of the notion of transversality for topological codes to include all local gates. 

\medskip
\noindent{\em Definitions---}
The notion of locality plays a crucial role here. For an operator $X$ acting on the qubits of a 2D lattice, let us denote by $|X|$ the range of $X$, defined as 
the size of the smallest square containing the support of $X$. With this definition, a Hamiltonian of the form \eq H is {\em local} if there exists a constant $w$ such that $|S_a| \leq w$ for all $a$. A 
translationally invariant
unitary transformation $U$ is {\em local} if there exists a constant $v$ such that $|U^\dagger XU| \leq |X| + v$ for all operator $X$. Note that this definition is equivalent \cite{ANW10a} to the requirement that $U$ be decomposable into a system-size independent sequence of nearest neighbor unitary transformations. Lastly, we will say that two local stabilizer codes defined by Hamiltonians $H$ and $H'$ \eq H, with stabilizer groups $\cS$ and $\cS'$, are {\em locally equivalent} if there exists a local unitary $U$ and two {\em trivial} LTI stabilizer groups $\cT$ and $\cT'$ such that $U(\cS\otimes \cT)U^\dagger = \cS'\otimes \cT'$. A trivial stabilizer group is generated only by single-qubit operators. Physically, $U$ takes the ground state of $H$ onto that of $H'$, and adds or removes extra qubits that are completely unentangled. The existence of renormalization group transformations that disentangle some qubit from topological codes \cite{AV08a} show the necessity of $\cT$ and $\cT'$ in this definition.

Kitaev's topological code \cite{Kit03a,DKLP02a} is defined on a 2D square lattice, with one qubit attached to each edge. 
For each lattice site $s$, define an operator $A_s = \prod_{e\in E_s} \sigma_z^e$ where $E_s$ denotes the set of edges incident to site $s$. Similarly, define for each lattice plaquette $p$ (site of the dual lattice) an operator $B_p = \prod_{e\in E_p} \sigma_x^e$ where $E_p$ denotes the set of edges adjacent to plaquette $p$. 
The Hamiltonian of the model is 
\be\label{KTC}
H = -\sum_s A_s - \sum_p B_p.
\ee
The excitations are anyons, gapped and topologically charged. Indeed, any set of excitations on the KTC can be reduced by local operations to one of four configurations:  the vacuum (0) corresponding to no excitations, an electric charge ($e$) corresponding to a plaquette excitation $B_p$, a magnetic charge ($m$) corresponding to a site excitation $A_s$, and a composite excitation ($f$) containing both. These four sectors are the topological charges of the model. Excitations with different charges are characterized by different topological interactions or braiding statistics. According to the effect of exchanging two identical charges, electric and magnetic particles are classified as bosons, while the composite particle is a fermion. As for mutual statistics, they are all semionic because braiding any two distinct non-vacuum charges yields a $-1$ phase. Finally, two charges can merge to form a new charge. The corresponding fusion rules are Abelian and such that $m\times e \rightarrow f$ and $\sigma \times \sigma \rightarrow 0$ for $\sigma = m,e,f$. The notion of topological charge is of utmost relevance because local
equivalence preserves the anyon model.

\medskip
\noindent{\em Main result---} 
We assume that 2D TSCs cannot give rise to chiral anyons \footnote{In Abelian models, the chiral central charge $c_-$ is related to the topological spins $\theta_i$ (=1 for bosons, -1 for fermions) through $ \sum_i \theta_i = \kappa e^{\pi i c_-/4}$ \cite{Kit06a}.} because the Hamiltonian terms $S_a$ commute with each other \cite{Kit06a}. Under this assumption, our main result is:
\begin{theorem}
Every 2D TSC is locally equivalent to a finite number of copies of KTC.
\end{theorem}
By $n$ copies of the code, we mean stacking $n$ lattices on top of each other, each with the same Hamiltonian \eqref{KTC}.
This result implies that equivalence classes are labeled by the total quantum dimension $\kappa$ of the code.
The proof is rather technical and we do not present it here. It has two main steps.
The first one shows that the excitations of a 2D TSC have the same topological charges as a number of copies of KTC.
Then, after a suitable renormalization, the hopping terms of the corresponding anyons are mapped to each other.

\BF
\includegraphics[width=9cm]{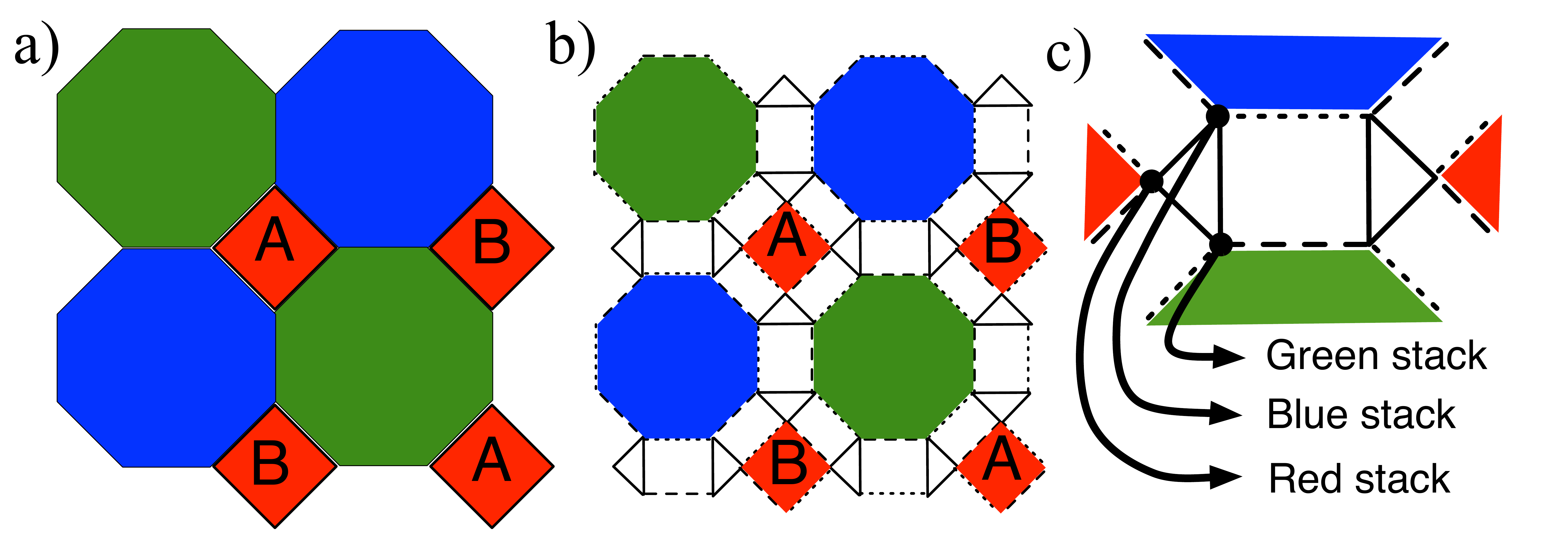}
\EF{a) Regular square-octagon lattice for TCC. The diamonds can be labeled {\sc A} or {\sc B} according to a chessboard pattern. There are two stabilizers \eq{genTCC} associated to each plaquette. b) Expanded square-octagon lattice for TSC. Starting with the lattice on the left, each vertex is expanded into a triangle. There is one gauge operator \eq{gaugeTSCC} per edge. c) Zoom of a region of the extended lattice and rearrangement of the qubit into three stacks.}{fig:48}

We illustrate this for topological color codes (TCCs) \cite{BM07a,B10b1}.
A TCC can be constructed on any 3-valent lattice with 3-colorable faces, but we take in particular the square-octagon regular lattice of \fig{48} a). This lattice is particularly useful in terms of fault-tolerance \cite{BM07a}. Qubits are located at the vertices of the lattice, and there are two stabilizer operators per plaquette $p$
\BE
S_p^\sigma = \bigotimes_{e\in E_p} \sigma^e, \quad {\rm with}\ \sigma \in \{\sigma_x,\sigma_z\}.
\label{eq:genTCC}
\EE 
The excitations in this model carry 16 different topological charges that correspond exactly to the charges obtained from two copies of KTC. Guided by this charge identification, we obtain the mapping shown at \fig{mapping}. It can be directly verified that it maps stabilizer generators of TCC to stabilizer generators of two KTC, in this case with no need to add trivial stabilizers.

\BF
	\includegraphics[width=8.5cm]{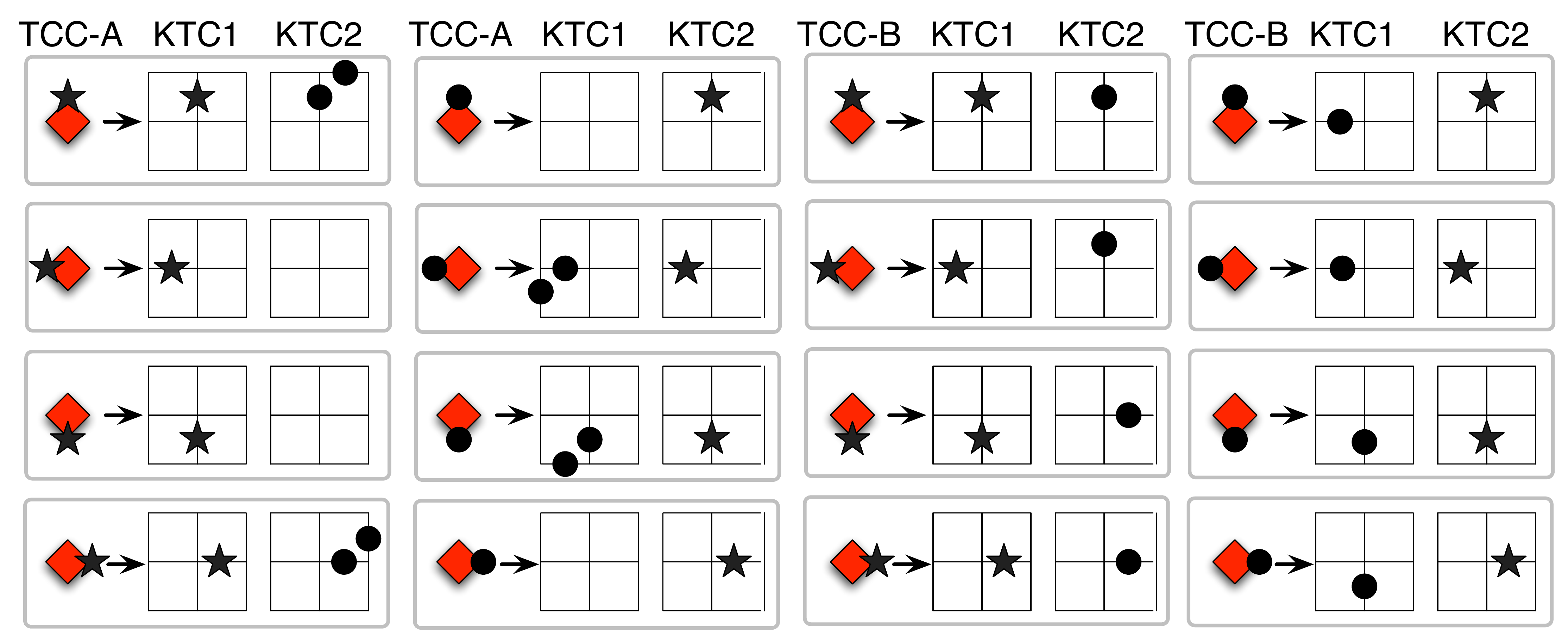}
\EF{
Mapping between the Pauli operators of the square-octagon TCC an two copies of Kitaev's code KTC1, KTC2. The first (last) two columns are for the $A$ ($B$) sub-lattice. Circles (stars) represent $\sigma_z$ ($\sigma_x$) operators. For instance, the upper left diagram indicates that a $\sigma_x$ located at the top of a diamond of the $A$ sub-lattice gets mapped to a $\sigma_x$ on KTC1 and two $\sigma_z$ on KTC2. All commutation relations are preserved by this mapping, so it is unitary and obviously local. }{fig:mapping}

\medskip
\noindent{\em Subsystem codes---}
Subsystem stabilizer codes form a more general class of stabilizer codes \cite{Pou05b,Bac05a}. They can be defined as a pair $(\cS,\cG)$, where  $\cG$ is an arbitrary Pauli subgroup and $\cS$ a stabilizer such that $\cS \propto \cZ(\cG) \cap \cG$. Encoding is not done on the whole subspace defined by $\cS$, but rather on the subsystem where the action of $\cG$ is trivial. This way, errors caused by operators in $\cG$ do not affect encoded states. Because of this, elements of $\cG$ are called gauge operators.

We say that a subsystem code $(\cS,\cG)$ is LTI if $\cG$ admits a LTI set of generators $\cG_b$. 
Note that some local subsystem codes admit no local stabilizer generators, e.g. \cite{Bac05a}.
Unlike them, a topological subsystem code should have a stabilizer with a local description. In addition, local operators should not recover any encoded information. Since we do not care about the effect of gauge operators, this can be formalized as follows in an infinite lattice:
\begin{definition}
A topological stabilizer subsystem code (TSSC) is a LTI subsystem stabilizer code $(\cS,\cG)$ such that
$\cZ(\cS)\propto \cG.$
\end{definition}

There is a general strategy to understand TSSCs in terms of TSCs. Namely, to find a TSC $\cS'$ that lies in between the stabilizer group and the gauge group of the subsystem code, i.e. $\cS \subset \cS' \subset \cG$. We can then map $\cS'$ invoking Theorem 1, which shows that $\cS$ is locally equivalent to a subset of the generators of several copies of KTCs. Not all topological charges of these KTCs carry information: some are associated to gauge operators. These ``gauge charges'' do not topologically interact with charges that describe errors in $\cS$, that we call ``proper charges''. Stabilizer of $\cS$ detect the presence of proper charges, ignoring any gauge charge. Unlike for TSCs, proper charges can give rise to a chiral anyon model as exemplified below.  

Let us illustrate this strategy with an important family of 2D subsystem codes \cite{B10b1} called topological subsystem color codes (TSCCs). Given the lattice of a TCC, we can inflate each vertex into a triangle as in \fig{48} b).  Qubits are located on the vertices of this inflated lattice, and there is one gauge group generator associated to each pair of sites $i,j$ connected by an edge
\BE
G_{ij} = \sigma^i\sigma^j
\label{eq:gaugeTSCC}
\EE 
with $\sigma = \sigma_x, \sigma_y$, or $\sigma_z$ for a dashed, dotted, or solid edges respectively.
This code admits a set of local stabilizer generators, some of which involve a relatively large number of qubits (up to 24). Errors are described by three nontrivial topological charges, all fermions ($f_1,f_2,f_3$), making it a chiral anyon model. All the mutual statistics are semionic and the fusion rules are $f\times f\rightarrow 0$ and $f_i \times f_j \rightarrow f_k$ with $i,j,k$ all different. These charges can be obtained as a subset of two copies of KTC. For instance, one can identify $f_1 \leftrightarrow [m,f]$, $f_2 \leftrightarrow [e,f]$, and $f_3 \leftrightarrow [f,0]$ where the notation $[a,b]$ stands for a composite particle of charge $a$ on the first KTC and charge $b$ on the second.
This suggests that we should be able to find a TSC $\cS'$ with  $\cS \subset \cS' \subset \cG$ and $\cS'$ locally equivalent to $n$ copies of KTC with $n\geq 2$.
We will present two different ways of obtaining $\cS'$ that are geometry independent (i.e., not restricted to the square-octagon lattice). 

In the first construction, $\cS'$ is the stabilizer of three TCC on the corresponding non-inflated lattice. Indeed, all we need to do is to rearrange the qubits. The three qubits located at the vertices of each triangle inherit the color label of the neighboring plaquette. We construct a stack of three TCC lattices---one per color---each one containing the qubits of that color, see \fig{48} c). It can be easily verified that this maps the generators of $\cS$ to a subset of the generators of the three TCCs. We obtain $\cS'$ by including the other generators of these TCCs. 
In the second construction, we consider the stabilizer group $\cS_z$ generated by the gauge operators of the form $\sigma_z\sigma_z$ (solid edges), which clearly is a subgroup of $\cZ(\cS)$. It follows from the results in \cite{BKM09a1} that $\cS':=\cS\cS_z$ is a TSC with the same topological charges as a TCC.
These two constructions illustrate that the quantum dimension of the intermediate code $\cS'$ is not uniquely determined, since in the first case we have $\kappa=2^6$ and in the second $\kappa=2^2$.

Because $\cS$ is a strict subset of $\cS'$, the mappings describe above do not take the system to the ground state of the resulting KTCs; the stabilizer added to $\cS$ to arrive at $\cS'$ will generally contain excitations. These can be eliminated by local transformations, fusing particle pairs into the vacuum. Moreover, because these excitations correspond to gauge charges, this local transformation does not change the encoded information. 

\BF
	\includegraphics[width=8.5cm]{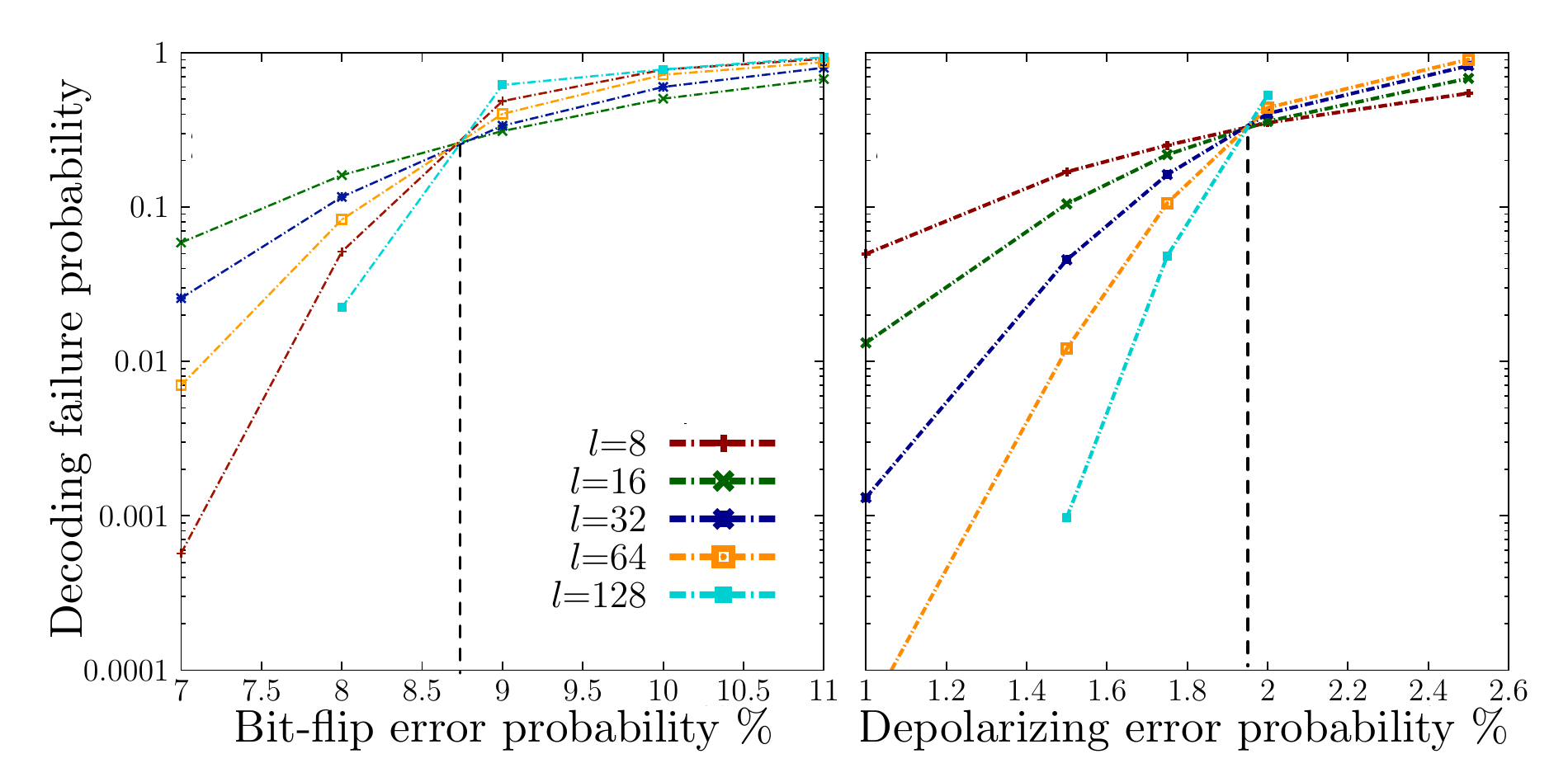}
\EF{Decoding failure probability as a function of the error probability of each qubit for the octagon-square TCC (left) and TSCC (right), based on the algorithm of \cite{DP10a}. The different curves illustrate lattices of different linear size $l$: below a threshold probability (dotted lines), the decoding failure probability decreases with the lattice size, leading to a perfect recovery in the thermodynamic limit. }{fig:48CCRes}

\medskip 
\noindent\textit{Decoding---}
When the system is prepared in the ground state of the Hamiltonian \eq H, all stabilizers have value $+1$. But in the presence of errors, this will not be the case in general. The problem of decoding a quantum code consists in identifying the most likely recovery to restore the encoded state from partial information coming from the measurement of the stabilizer operators, whose $\pm 1$ outcomes are called error syndrome.
Not all codes can be decoded efficiently, but fast approximate algorithms have been devised for KTC. 
The one presented in \cite{DKLP02a} uses Edmonds' minimum matching algorithm \cite{E65a} to find the shortest path that recombines all electric particles in pairs and independently all magnetic particles in pairs. For $N$ qbits, it runs in time $N^3$. The algorithm proposed in \cite{DP10a} uses renormalization group approximations to find the homological class of errors with the highest probability. It runs in time $\log N$. An efficient decoder was also devised for a TSSC on a particular lattice \cite{SBT10a}, but in general each new code requires a tailored decoding technique.

Our techniques can be used to decode any 2D TSCs or TSCCs. The idea is to choose a set of ``elementary'' charges---either fermionic or bosonic---that generate all the topological charges in the code. Then, the decoding problem can be mapped to that of KTC described above for each elementary charge. We have used this technique, combined to the decoding algorithm of \cite{DP10a},  for the TCC on the square-octagon lattice of \fig{48} on a bit-flip channel and found an error threshold of roughly $8.7$\%, in good agreement with the Monte Carlo estimate of $10.9$\% \cite{KBAM10a}
for ideal error correction. We have also used this technique for the TSCC on the square-octagon lattice of \fig{48} on a depolarizing channel and found an error threshold of roughly $1.95$\%, in good agreement with the estimate of $2$\% \cite{SBT10a} for a closely related code in a five-square lattice. 

\noindent\textit{Conclusion}---We have demonstrated that 2D topological stabilizer codes all belong to one universal topological phase by constructing an explicit local mapping onto multiple copies of Kitaev's topological code. This results also carries to a certain class of 2D subsystem codes, and in particular to all topological subsystem color codes. These local maps enable us to extend many properties of Kitaev's code to all 2D codes, and in particular directly yield efficient decoding algorithms for error correction. It could also have important implications for fault-tolerant quantum computation.

\noindent\textit{Acknowledgements---}
We thank Sergey Bravyi for stimulating discussions. This work was partially funded by NSERC, FQRNT, and MITACS.


\end{document}